\documentclass[reprint,
superscriptaddress,
amsmath,amssymb,
aps,
prl,
]{revtex4-1}

\usepackage{graphicx}
\usepackage{comment}
\usepackage{dcolumn}
\usepackage{color}
\usepackage{bm}
\usepackage{url}
\usepackage{braket}
\usepackage{amsfonts,amsmath,mathtools}
\usepackage{dsfont}
\usepackage{hyperref}
\usepackage[all]{hypcap}
\setcitestyle{numbers,square}

\newcommand*{\defeq}{\mathrel{\vcenter{\baselineskip0.5ex\lineskiplimit0pt\hbox{\scriptsize.}\hbox{\scriptsize.}}}=}

\begin{document}

\title{Eigenstate Thermalization in a Locally Perturbed Integrable System}

\author{Marlon Brenes}
\email{Corresponding author: brenesnm@tcd.ie}
\affiliation{School of Physics, Trinity College Dublin, College Green, Dublin 2, Ireland}
\author{Tyler LeBlond}
\affiliation{Department of Physics, The Pennsylvania State University, University Park, PA 16802, USA}
\author{John Goold}
\affiliation{School of Physics, Trinity College Dublin, College Green, Dublin 2, Ireland}
\author{Marcos Rigol}
\affiliation{Department of Physics, The Pennsylvania State University, University Park, Pennsylvania 16802, USA}

\begin{abstract}
Eigenstate thermalization is widely accepted as the mechanism behind thermalization in generic isolated quantum systems. Using the example of a single magnetic defect embedded in the integrable spin-1/2 $XXZ$ chain, we show that locally perturbing an integrable system can give rise to eigenstate thermalization. Unique to such setups is the fact that thermodynamic and transport properties of the unperturbed integrable chain emerge in properties of the eigenstates of the perturbed (nonintegrable) one. Specifically, we show that the diagonal matrix elements of observables in the perturbed eigenstates follow the microcanonical predictions for the integrable model, and that the ballistic character of spin transport in the integrable model is manifest in the behavior of the off-diagonal matrix elements of the current operator in the perturbed eigenstates.
\end{abstract}

\maketitle

How do statistical ensembles and thermal behavior emerge from the fundamental unitary dynamics of isolated quantum systems? This question, first posed in the earliest days of quantum mechanics~\cite{Schrodinger:1927, Vonneumann:1929, Goldstein:2010}, is still at the forefront of modern research in quantum statistical mechanics~\cite{Alessio:2016, Eisert:2015, Polkovnikov:2011}. The current interest in this foundational topic can be attributed to advances in ultracold atomic experiments where many-body systems can be time propagated coherently over unprecedented time scales~\cite{langen2015ultracold, Bloch:2012, Lewenstein:2007}. In particular, seminal experiments have demonstrated that integrability inhibits thermalization~\cite{Kinoshita:2006}, and that integrability breaking perturbations can be used to controllably bring a system to thermal equilibrium~\cite{tang_kao_18}.

The latter experimental results are consistent with the expectation that generic isolated quantum systems thermalize to a microcanonical distribution consistent with their energy density. The accepted mechanism for this is eigenstate thermalization, as prescribed by the eigenstate thermalization hypothesis (ETH)~\cite{Deutsch:1991, Srednicki:1994, Srednicki:1999, Rigol:2008, Alessio:2016}. For an observable $\hat O$, the ETH for the matrix elements $O_{nm}=\langle n|\hat O|m\rangle$ in the energy eigenbasis ($\hat H|m\rangle=E_m|m\rangle$) reads
\begin{equation}
\label{eq:eth}
O_{n m} = O(\bar{E}) \delta_{n m} + e^{-S(\bar{E}) / 2}f_{O}(\bar{E}, \omega)R_{n m},
\end{equation}
where $\bar{E} \defeq(E_{n} + E_{m}) / 2$ and $\omega \defeq E_{m} - E_{n}$. $S(\bar{E})$ is the thermodynamic entropy at energy $\bar{E}$, $R_{n m}$ is a random variable with zero mean and unit variance, and $O(\bar{E})$ and $f_{O}(\bar{E}, \omega)$ are smooth functions. The first term in Eq.~\eqref{eq:eth} advances that the diagonal matrix elements of observables are smooth functions of the energy $E_n$ (the eigenstate to eigenstate fluctuations are exponentially small in the size of the system~\cite{steinigeweg:2013, Kim_Huse_14, beugeling2014finite, Mondaini:2016, yoshizawa2018numerical, Vidmar_Fabian_19, Leblond:2019}). From the second term we see that the off-diagonal matrix elements are exponentially small in the system size (because of $e^{-S(\bar{E}) / 2}$) and that, up to random fluctuations, they are characterized by smooth functions $f_{O}(\bar{E}, \omega)$~\cite{Khatami:2013, Moessner:2015, Alessio:2016, Mondaini:2017, Vidmar_Fabian_19, Leblond:2019}. Those functions carry important information on fluctuation dissipation relations~\cite{Srednicki:1999, Khatami:2013, Alessio:2016}, and even on the multipartite entanglement structure of the energy eigenstates~\cite{Brenes2}.

Integrable systems, which possess extensive sets of nontrivial conserved quantities, do not follow the ETH. The diagonal matrix elements of observables exhibit eigenstate to eigenstate fluctuations that do not vanish in the thermodynamic limit~\cite{Rigol:2008, rigol2009breakdown, *rigol_offd_int1, rigol_offd_int2, steinigeweg:2013, beugeling2014finite, vidmar2016, Leblond:2019}, while their variance vanishes as a power law in the system size~\cite{biroli2010effect, ikeda2013finite, alba2015, Leblond:2019}. Because of this, in general, integrable systems do not thermalize~\cite{rigol_16}. They do equilibrate and, after equilibration, they are described by generalized Gibbs ensembles (GGEs)~\cite{rigol_dunjko_07, vidmar2016, essler_fagotti_review_16, caux_review_review_16}. For the off-diagonal matrix elements of observables in interacting integrable systems, it was recently shown that their variance is a well-defined (exponentially small in the system size) function of the average energy and the energy difference of the eigenstates involved~\cite{mallayya, Leblond:2019}, like in systems that satisfy the ETH. 

Integrability is believed to be unstable to perturbations~\cite{Alessio:2016}. Surprisingly, it has been shown that even a single magnetic impurity perturbation at the center of the integrable spin-1/2 $XXZ$ chain is enough to induce level repulsion and random matrix statistics in the spectrum~\cite{Santos:2004, santos2011domain, torres2014local, Torres_Herrera_2015, XotosIncoherentSIXXZ, Metavitsiadis1, Brenes:2018}. Recently, a study of both linear response and steady-state transport showed that this model displays ballistic spin transport~\cite{Brenes:2018}, challenging our expectation that quantum chaotic systems (those exhibiting random matrix statistics in the spectrum) should exhibit diffusive transport. In this Letter we show that the matrix elements of observables in such a model are fully consistent with the ETH. Unique to breaking integrability with local perturbations, we argue that statistical mechanics and transport properties of the unperturbed integrable model can end up embedded in properties of the eigenstates of the perturbed (quantum chaotic) one.

The Hamiltonian of the spin-1/2 $XXZ$ (in short, the $XXZ$) chain  can be written as (we set $\hbar = 1$):
\begin{equation}
\label{eq:h_xxz}
\hat{H}_{XXZ} = \sum_{i=1}^{N-1}\left(\hat{\sigma}^x_{i}\hat{\sigma}^x_{i+1} + \hat{\sigma}^y_{i}\hat{\sigma}^y_{i+1} + \Delta\,\hat{\sigma}^z_{i}\hat{\sigma}^z_{i+1}\right),
\end{equation} 
where $\hat{\sigma}^\nu_{i}$, $\nu = x,y,z$, correspond to Pauli matrices in the $\nu$ direction at site $i$ in a chain with $N$ (taken to be even) sites and open boundary conditions. In Eq.~\eqref{eq:h_xxz}, $\Delta$ is the anisotropy parameter. We focus on $\Delta = 0.55$, for which spin transport is ballistic, but also show results for $\Delta = 1.1$, for which spin transport is diffusive~\cite{2020arXiv200303334B}.

The $XXZ$ chain is a quintessential interacting integrable model~\cite{ShastryBethe1990, Cazalilla:2011}. We study properties of its eigenstates along with those of eigenstates of the nonintegrable model obtained by perturbing it with a magnetic impurity about the center of the chain. This local perturbation produces an energy spectrum with a Wigner-Dyson distribution of nearest-neighbor level spacings \cite{Santos:2004, santos2011domain, torres2014local, XotosIncoherentSIXXZ, Metavitsiadis1, Brenes:2018}. The single-impurity Hamiltonian has the form
\begin{equation}
\label{eq:h_si}
\hat{H}_{\textrm{SI}} = \hat{H}_{XXZ} + h\, \hat{\sigma}^z_{N/2},
\end{equation}
where $h$ is the strength of the magnetic impurity. We henceforth set $h = 1$ so that all energy scales in our perturbed Hamiltonian are $\mathcal{O}(1)$. 

Both Hamiltonians of interest in this work, Eqs.~\eqref{eq:h_xxz} and~\eqref{eq:h_si}, commute with the total magnetization operator in the $z$ direction, $[\hat{H}_{XXZ}, \sum_i\hat{\sigma}^z_i]=[\hat{H}_{\textrm{SI}}, \sum_i\hat{\sigma}^z_i]=0$, so they are $U(1)$ symmetric. We focus on the zero magnetization sector, $\sum_i \braket{\hat{\sigma}^z_i} = 0$, which is the largest sector. Reflection symmetry is present in $\hat{H}_{XXZ}$. We explicitly break it by adding a very weak magnetic field at site $i=1$, $h_1=10^{-1}$  (like open boundary conditions, this perturbation does not break integrability~\cite{Santos:2004}). We use state of the art full exact diagonalization, and study chains with up to $N = 20$ sites, for which the Hilbert space dimension $\mathcal{D} = N! / [(N/2)!]^2 = 184\,756$.

\emph{Diagonal ETH}.--- Let us first study the diagonal matrix elements of two related local observables. We choose the local kinetic energy at site $i = N/4$ (far away from the boundary and the impurity), 
\begin{equation}
\label{eq:kobs}
\hat{K} \defeq \hat{K}_{\frac{N}{4}, \frac{N}{4}+1} = \left( \hat{\sigma}^x_{\frac{N}{4}}\hat{\sigma}^x_{\frac{N}{4}+1} + \hat{\sigma}^y_{\frac{N}{4}}\hat{\sigma}^y_{\frac{N}{4} + 1} \right),
\end{equation}
and the total kinetic energy per site, the average local kinetic energy, defined as
\begin{equation}
\label{eq:tobs}
\hat{T} \defeq \frac{1}{N} \sum_{i=1}^{N-1} \left( \hat{\sigma}^x_{i}\hat{\sigma}^x_{i+1} + \hat{\sigma}^y_{i}\hat{\sigma}^y_{i+1} \right).
\end{equation}
The contrast between the two shows the effect of averaging in nontranslation invariant systems. Qualitatively similar results were obtained for other local observables.

\begin{figure}[!t]
\includegraphics[width=\columnwidth]{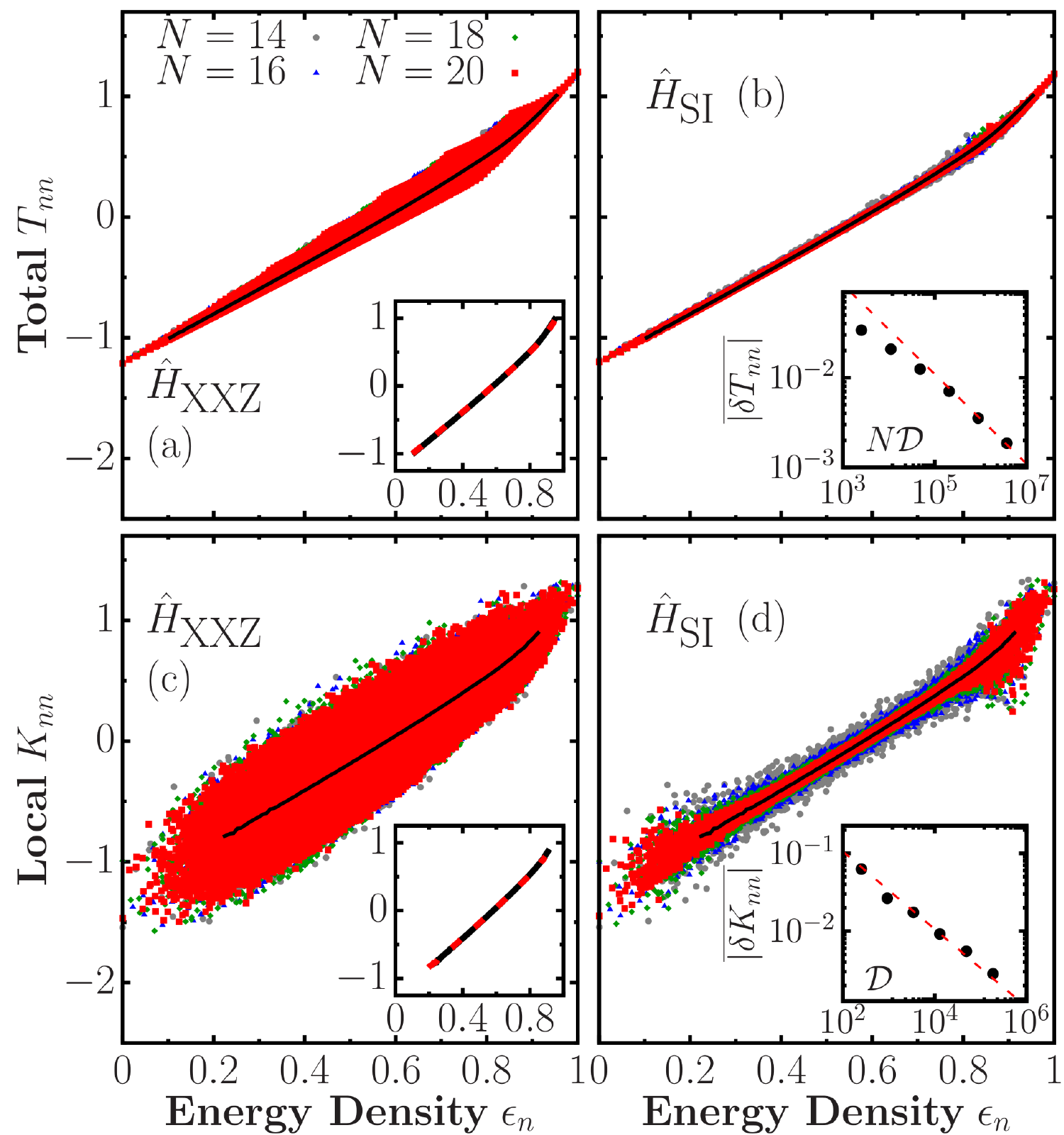}
 \caption{Diagonal matrix elements of $\hat T$ [(a), (b)] and $\hat K$ [(c), (d)] in the eigenstates of $\hat{H}_{XXZ}$ [(a), (c)] and $\hat{H}_{\textrm{SI}}$ [(b), (d)] ($\Delta=0.55$). The black lines show microcanonical averages (within windows with $\delta \epsilon_n = 0.008$) in $\hat{H}_{XXZ}$ for the largest chain ($N = 20$). The insets in (a) and (c) show the equivalence of the microcanonical predictions in both models for each observable, while the insets in (b) and (d) show the $(N\mathcal{D})^{-1/2}$ and $\mathcal{D}^{-1/2}$ scalings, respectively, of $\overline{|\delta O_{nn}|} \defeq \overline{|O_{nn} - O_{n+1n+1}|}$ (the dashed lines are $\propto x^{-1/2}$), where we average over the central 20\% of the eigenstates in chains with $N = 10$, 12, $\dots$, 20.}
\label{fig:1}
\end{figure}

In Fig.~\ref{fig:1}, we show the diagonal matrix elements of $\hat K$ and $\hat T$ in the eigenstates of the Hamiltonians in Eqs.~\eqref{eq:h_xxz} and \eqref{eq:h_si}. The results are plotted as functions of the energy density defined as $\epsilon_{n} \defeq E_{n} - E_{\textrm{min}} / E_{\textrm{max}} - E_{\textrm{min}}$, where $E_{n}$ is the $n$th energy eigenvalue, and $E_{\textrm{min}}$ ($E_{\textrm{max}}$) is the lowest (highest) energy eigenvalue. Despite the quantitative differences in the behavior of the two observables in each model (at each energy, the spread of $T_{nn}$ is smaller than that of $K_{nn}$), they both exhibit a qualitatively similar behavior depending on whether the model is integrable ($\hat{H}_{XXZ}$) or nonintegrable ($\hat{H}_{\textrm{SI}}$). In the integrable model, the spread of $T_{nn}$ and $K_{nn}$ at each energy does not change with changing system size (the system does not satisfy the ETH), while in the nonintegrable model it decreases exponentially fast with increasing $N$ [away from the edges of the spectrum, see insets in Figs.~\ref{fig:1}(b) and~\ref{fig:1}(d) for a variance indicator] suggesting that $T_{nn}$ and $K_{nn}$ satisfy the ETH~\cite{torres2014local, Torres_Herrera_2015}.

Since the single impurity is a subextensive local perturbation to the $XXZ$ chain, it does not affect the microcanonical predictions (away from the edges of the spectrum) for local observables (away from the impurity) in sufficiently large system sizes. This is confirmed in the insets in Figs.~\ref{fig:1}(a) and~\ref{fig:1}(c). Hence, a remarkable consequence of the single impurity producing eigenstate thermalization (something that is achieved via mixing nearby unperturbed energy eigenstates) is that the smooth functions $T_{nn}$ and $K_{nn}$ are nothing but the microcanonical ensemble predictions for the integrable model. Another interesting consequence of it is that if one evolves highly excited eigenstates of $\hat{H}_{\textrm{SI}}$ under the integrable dynamics generated by $\hat{H}_{XXZ}$, thermalization will occur at long times (as in the limit of vanishingly small but extensive integrability breaking perturbations~\cite{rigol_16, rigol_srednicki_12}).

\emph{Off-diagonal ETH}.--- Next we study the off-diagonal matrix elements of the total kinetic energy per site $\hat T$ [Eq.~\eqref{eq:tobs}], and of the spin current operator per site $\hat J$,
\begin{equation}
\label{eq:jobs}
\hat{J} \defeq \frac{1}{N} \sum_{i=1}^{N-1} \left( \hat{\sigma}^x_{i}\hat{\sigma}^y_{i+1} - \hat{\sigma}^y_{i}\hat{\sigma}^x_{i+1} \right).
\end{equation}

Since $\hat T$ and $\hat J$ have Hilbert-Schmidt norms that scale as $1/\sqrt{N}$, the off-diagonal part of the ETH needs to be modified to read~\cite{Leblond:2019, vidmar2020}  
\begin{equation}
\label{eq:eth_scaled}
O_{nm} = \frac{e^{-S(\bar{E})/2}}{\sqrt{N}}f_O(\bar{E}, \omega)R_{nm}.
\end{equation}
We focus on the ``infinite-temperature'' regime, in which $\bar{E}\approx 0$ and $S(\bar{E}) \approx \ln{\mathcal{D}}$. 

\begin{figure}[!t]
\includegraphics[width=\columnwidth]{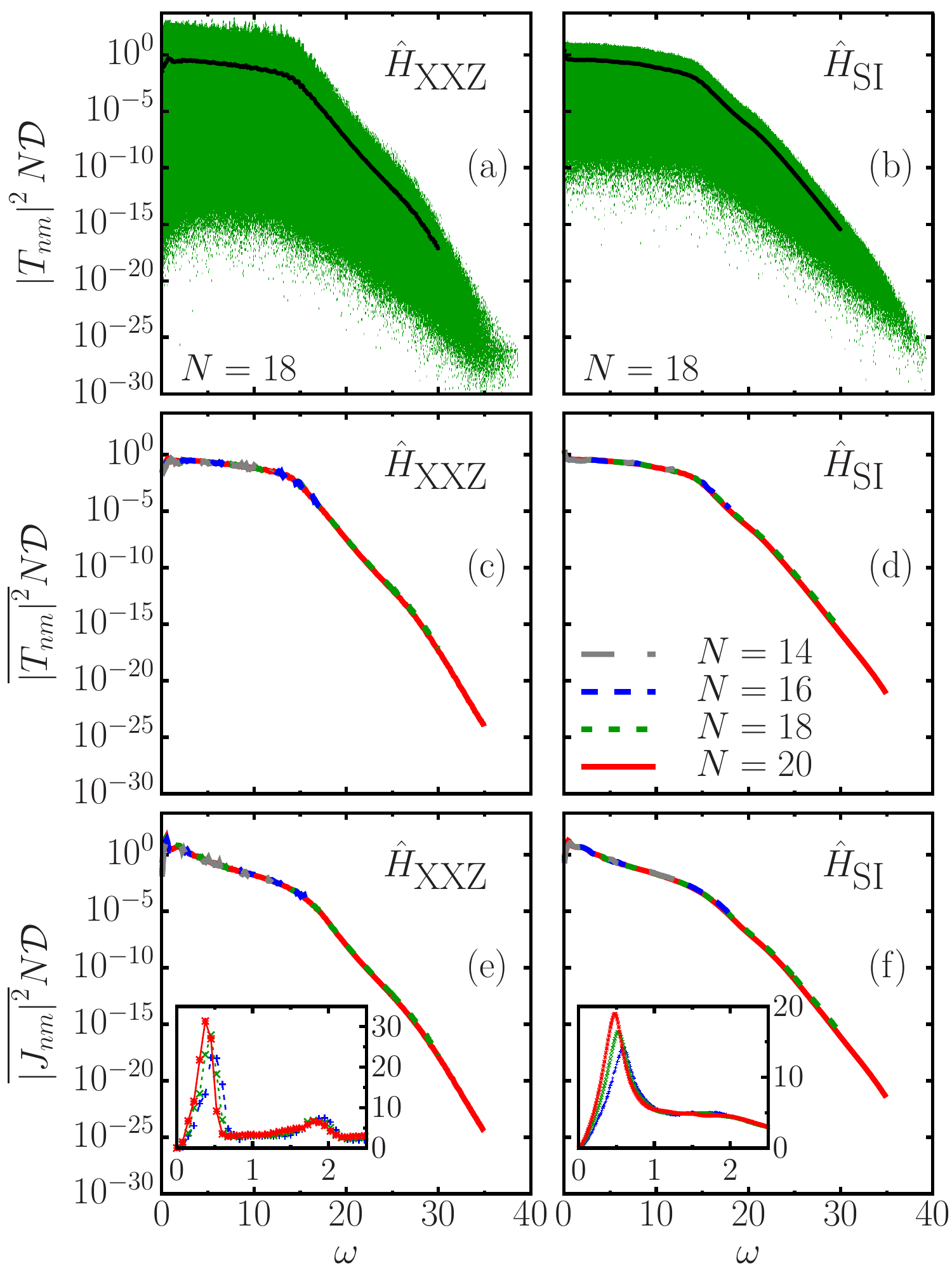}
 \caption{[(a), (b)] Off-diagonal matrix elements of $\hat T$, and the corresponding coarse-grained average [continuous (black) line], plotted vs $\omega$ for chains with $N = 18$. [(c), (d)] Coarse-grained averages of $T_{nm}$, including the ones reported in (a) and (b), for different chain sizes. [(e), (f)] Coarse-grained averages of $ J_{nm}$ for different chain sizes (the insets show results at low $\omega$, see also Fig.~\ref{fig:4}). The left panels [(a), (c), and (e)] show results for $\hat{H}_{XXZ}$, while the right ones [(b), (d), and (f)] show results for $\hat{H}_{\textrm{SI}}$ ($\Delta=0.55$). The matrix elements were computed within a small window of energy around $\bar{E} \approx 0$ (center of the spectrum) of width $0.05\varepsilon$ ($0.075\varepsilon$ for the insets), where $\varepsilon \defeq E_{\textrm{max}} - E_{\textrm{min}}$. The coarse-grained averages were computed using a window $\delta \omega = 0.1$ [$\delta \omega = 0.075$ and $\delta \omega = 0.01$ for the insets in (e) and (f), respectively].}
\label{fig:2}
\end{figure}

In Figs.~\ref{fig:2}(a) and~\ref{fig:2}(b), we show the off-diagonal matrix elements $|T_{nm}|^2$ in the $XXZ$ and single-impurity models, respectively. As expected, their overall dispersion is larger in the former (integrable) model than the latter (nonintegrable) one. For both models, Figs.~\ref{fig:2}(a) and~\ref{fig:2}(b) show that the coarse-grained average $\overline{|T_{nm}|^2}$ (which corresponds to the variance of the off-diagonal matrix elements as $\overline{T_{nm}}=0$) is a smooth function of $\omega$~\cite{Leblond:2019}. In Ref.~\cite{Leblond:2019}, it was shown that the variance of the off-diagonal matrix elements of observables like the ones of interest here satisfies $\overline{|O_{nm}|^2} \propto (N\mathcal{D})^{-1}$ both for integrable interacting and nonintegrable models. Figures~\ref{fig:2}(c) and~\ref{fig:2}(d) for $\overline{|T_{nm}|^2}$, and Figs.~\ref{fig:2}(e) and~\ref{fig:2}(f) for $\overline{|J_{nm}|^2}$, show that such a scaling is satisfied by our observables in the $XXZ$ and single-impurity models.

\begin{figure}[!t]
\includegraphics[width=0.9\columnwidth]{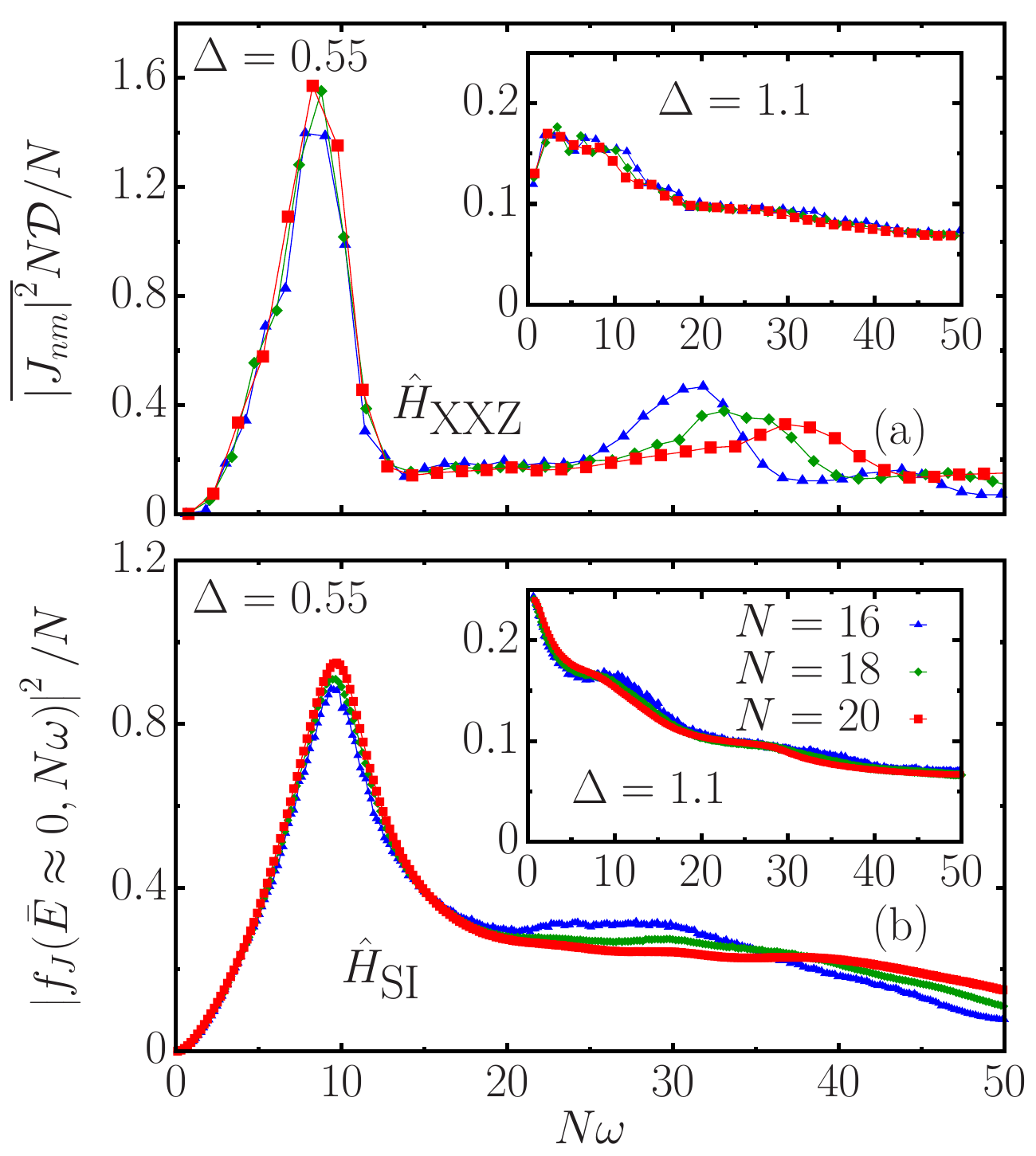}
\caption{Scaled variances of the off-diagonal matrix elements of $\hat J$ in the eigenstates of $\hat{H}_{XXZ}$ (a) and $\hat{H}_{\textrm{SI}}$ (b) plotted vs $N\omega$. The main panels (insets) show results for $\Delta=0.55$ ($\Delta=1.1$). The matrix elements were computed within a small window of energies around $\bar{E} \approx 0$ of width $0.075\varepsilon$. For the binned averages, we used $\delta \omega = 0.075$ in (a) and $\delta \omega = 0.01$ in (b).}
\label{fig:4}
\end{figure}

Figures~\ref{fig:2}(c) and~\ref{fig:2}(d) [Figs.~\ref{fig:2}(e) and~\ref{fig:2}(f)] also show that the variances $\overline{|T_{nm}|^2}$ ($\overline{|J_{nm}|^2}$) are very similar in the two models (the differences are consistent within present finite-size effects). For $\overline{|J_{nm}|^2}$, see insets in Figs.~\ref{fig:2}(e) and~\ref{fig:2}(f), the similarity extends to features that occur at low frequencies (see also Fig.~\ref{fig:4}). This opens the question of whether there is any difference between the off-diagonal matrix elements of observables in both models.

We find that the off-diagonal matrix elements of observables are normally distributed in the single-impurity model (qualitatively similar results have been obtained in other nonintegrable models~\cite{Moessner:2015, Mondaini:2017, Leblond:2019}), while they are close to log-normally distributed in the $XXZ$ chain~\cite{Leblond:2019}. To test the normality of the distribution in the single-impurity model for different values of $\omega$, and to contrast it to the results for the $XXZ$ chain, we compute~\cite{Leblond:2019}
\begin{equation}
\label{eq:gamma}
\Gamma_{\hat{O}}(\omega) \defeq \overline{|O_{nm}|^2} / \overline{|O_{nm}|}^2.
\end{equation}
$\Gamma_{\hat{O}} = \pi / 2$ for normally distributed matrix elements.

\begin{figure}[!t]
\includegraphics[width=\columnwidth]{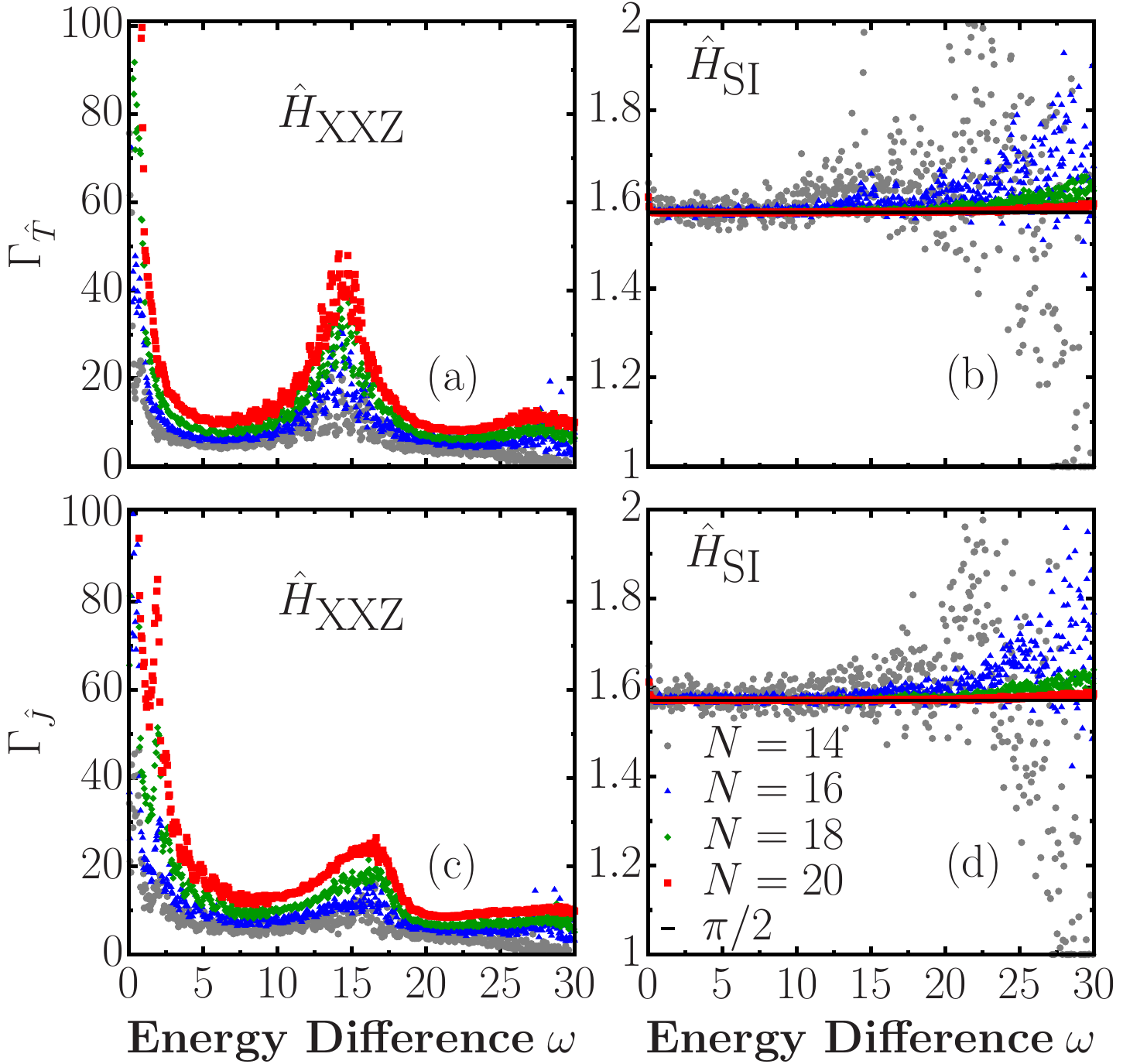}
\caption{$\Gamma_{\hat{O}}(\omega)$, see Eq.~\eqref{eq:gamma}, for the total kinetic energy per site [(a), (b)] and for the current operator [(c), (d)], in the $XXZ$ [(a), (c)] and single-impurity [(b), (d)] models ($\Delta=0.55$). The horizontal line in (b) and (d) marks $\pi/2$. The matrix elements were computed using the same energy window as in Fig.~\ref{fig:2}, while the coarse-graining parameter is $\delta \omega = 0.05$.}
\label{fig:3}
\end{figure}

In Fig.~\ref{fig:3}, we show results for $\Gamma_{\hat{T}}(\omega)$ [Figs.~\ref{fig:3}(a) and~\ref{fig:3}(b)] and $\Gamma_{\hat{J}}(\omega)$ [Figs.~\ref{fig:3}(c) and~\ref{fig:3}(d)] in the $XXZ$ [Figs.~\ref{fig:3}(a) and~\ref{fig:3}(c)] and single-impurity [Figs.~\ref{fig:3}(b) and~\ref{fig:3}(d)] models. For all values of $\omega$ shown in Figs.~\ref{fig:3}(b) and~\ref{fig:3}(d) for the single-impurity model, $\Gamma_{\hat T}(\omega)$ and $\Gamma_{\hat J}(\omega)$, respectively, approach $\pi / 2$ as $N$ increases, i.e., $T_{nm}$ and $J_{nm}$ are well described by a normal distribution. On the other hand, in Figs.~\ref{fig:3}(a) and~\ref{fig:3}(c) for the $XXZ$ model, $\Gamma_{\hat T}(\omega)$ and $\Gamma_{\hat J}(\omega)$, respectively, depend on the system size, i.e., $T_{nm}$ and $J_{nm}$ are not normally distributed.

The results discussed so far for the matrix elements of local operators in the single-impurity model show that they are fully consistent with the ETH. The fact that the off-diagonal matrix elements are normally distributed (the variance sets all central moments) means that one can define a meaningful $f_{O}(\bar{E}, \omega)$, while this is not the case for the $XXZ$ chain. The question we address next is related to the ballistic spin transport in the single-impurity model~\cite{Brenes:2018}, which is in stark contrast to the usual diffusive transport found in nonintegrable models.

\emph{Ballistic transport}.--- Within linear response, the real part of the conductivity reads ($k_B = 1$) \cite{kubo1957statistical, kubo1957statistical2, ShastryKubo2008, RigolDrude:2008, 2020arXiv200303334B}
\begin{align}
\label{eq:kuboformula}
\textrm{Re}[\sigma_N(\omega)] &= \pi D_N\delta(\omega) + \\
\frac{\pi}{N}&\left(\frac{1-e^{-\beta \omega}}{\omega}\right)\sum_{\epsilon_n \neq \epsilon_m} p_n|J_{nm}|^2\delta(\epsilon_m - \epsilon_n - \omega),\nonumber 
\end{align}
where $D_N$ is known as the Drude weight, $\beta$ is the inverse temperature, $p_n = e^{-\beta E_n} / Z$ is the Boltzmann weight of eigenstate $\ket{n}$, and $Z$ is the partition function. $J_{nm}$ are the matrix elements of the spin current operator. In integrable systems with open boundary conditions (e.g., our $XXZ$ chain), $D_N$ can be proved to be identically zero no matter the nature of the spin transport~\cite{RigolDrude:2008}. When transport is ballistic, a peak (or peaks) appear in $\textrm{Re}[\sigma_N(\omega)]$ at a nonzero frequency (frequencies) proportional to $1/N$. When $N\rightarrow\infty$, the peak (peaks) move toward $\omega\rightarrow0$ resulting in a peak in $\textrm{Re}[\sigma_N(\omega=0)]$ that signals ballistic transport~\cite{RigolDrude:2008}. Exactly the same was shown to occur in our single impurity model in Ref.~\cite{Brenes:2018}. Therefore, in our integrable and nonintegrable models ballistic transport emerges because of the $\omega\rightarrow 0$ behavior of the off-diagonal matrix elements of the current operator. 

In Fig.~\ref{fig:4}(a), we show the scaled variances of the matrix elements of $\hat J$ in $XXZ$ chains with $N=16$, 18, and 20 as functions of $N\omega$ for $\Delta=0.55$. A large peak can be seen at a frequency that scales as $1/N$ whose area does not change with increasing $N$. This is consistent with the behavior of $\textrm{Re} [\sigma_N(\omega)]$~\cite{RigolDrude:2008, Brenes:2018} signaling coherent transport~\cite{ZnidaricXXZspintransport}. The position of the smaller (second) peak is nearly $N$ independent [see inset in Fig.~\ref{fig:2}(e)], appearing to mark the onset of the $N$-independent behavior shown in Fig.~\ref{fig:2}. The variances of the matrix elements of $\hat J$ in the (nonintegrable) single-impurity model, which, remarkably, define a novel $N$-independent ETH function $|f_{J}(\bar{E}\approx 0, N\omega)|^2/N$ [Fig.~\ref{fig:4}(b)], display the same low-frequency behavior as in the (integrable) $XXZ$ chain. In contrast, as shown in the insets in Fig.~\ref{fig:4}, the scaled variances of the matrix elements of $\hat J$ behave completely differently for $\Delta=1.1$ (for which spin transport is diffusive). The nature of the spin transport in the absence and presence of the single magnetic defect, for $\Delta$ in the easy-plane and easy-axis regimes, is something that can readily be probed in ultracold gases experiments~\cite{2020arXiv200509549J}.
 
\emph{Conclusions}.--- We showed that the ETH is fully fulfilled when breaking integrability with a local perturbation and that, in such setups, it can inherit statistical mechanics and transport properties of the integrable model. Specifically, we showed that the diagonal matrix elements of observables in the perturbed energy eigenstates can follow the microcanonical predictions for the integrable model, and that ballistic transport in the integrable model can result in a novel $N$-independent ETH function $|f_{J}(\bar{E}\approx 0, N\omega)|^2/N$ that characterizes the off-diagonal matrix elements of the current operator in the perturbed energy eigenstates at low frequencies.

\emph{Acknowledgements}.--- This work was supported by the European Research Council Starting Grant ODYSSEY Grant No.~758403 (M.B.~and J.G.), the Royal Society (M.B.), a SFI-Royal Society University Research Fellowship (J.G.), and the National Science Foundation Grant No.~PHY-1707482 (T.L. and M.R.). M.B.~and J.G.~acknowledge the DJEI/DES/SFI/HEA Irish Centre for High-End Computing (ICHEC) for the provision of computational facilities and support, project TCPHY118B, and the Trinity Centre for High-Performance Computing. 

\emph{Note added}.--- Recently, other works have appeared exploring the structure of the off-diagonal matrix elements of different classes of observables in integrable and locally perturbed integrable models such as the ones considered here~\cite{polkovnikovprep2020, brenesprep2020, santos2020speck}, and in central spin models~\cite{anushyaprep2020}.

\bibliography{bibliography.bib}

\end{document}